# Telescopic system with a rotating objective element

G. Chadzitaskos and J. Tolar

Faculty of Nuclear Sciences and Physical Engineering

Czech Technical University

Brehova 7

CZ - 115 19 Prague 1

Czech Republic

Abstract

  The angular resolution is the ability of a telescope to render detail: the higher the resolution the finer is the detail. It is, together with the aperture, the most important characteristic of telescopes.

  We propose a new construction of telescopes with improved ratio of angular resolution and area of the primary optical element (mirror or lense). For this purpose we use the rotation of the primary optical element with one dominating dimension. The length of the dominating dimension of the primary optical element determines the angular resolution. During the rotation a sequence of images is stored in a computer and the images of observed objects can be reconstructed using a relatively simple software. The angular resolution is determined by the maximal length of the primary optical element of the system. This construction of telescopic systems allows to construct telescopes of high resolution with lower weight and fraction of usual costs.

Introduction

Up to now telescopes are used with objectives of circular or regular polygonal form. The angular resolution of such telescopes is almost the same in all lateral directions. As a rule, a telescope is supported by a mount which moves the telescope to point its optical axis to the object to be observed. In the case of moving objects, the mount makes the optical axis follow the position of the object. There are three types of mountings: paralactical, azimuthal and four axes mountings. The azimuthal mount has horizontal and vertical rotational axes and it is necessary to perform both rotations simultaneusly during the observations of moving objects. The paralactical mount has two perpendicular axes, one axis is parallel with the Earth axis, and a suitable rotation around this axis makes the telescope follow the celestial objects. The four axes mounting is useful for observation of moving objects like satellites. Three axes are oriented in such directions that the pointing of the object is done by a suitable rotation around the fourth axis. Most contemporary mountings are controlled by computers. To point the object it is enough to input the coordinates of the object, or sometimes the name or the number of the object into computer, and all control is performed automatically.

All known telescopic systems are of two types: the Galileo and Kepler with a lense (or several optical elements acting as a lense with corrections of the optical defects) as the primary optical element (refractor), or the Newton and Cassegrain with a parabolical mirror (reflector) as the primary optical element. On the output end of a telescope an eyepiece, photographic camera or CCD camera is placed to view the image, or a light spectrometer. The output from CCD or spectrometer is stored in a computer for further elaboration.

The refractor uses an objective lense that bends light rays for delivery to the eyepiece. The parabolic mirror of the reflector collects the light on a secondary mirror and the secondary mirror reflects light - in the Newton case into the eyepiece on the side of the telescope, in the Cassegrain case light passes through the hole in the primary mirror.

Angular resolution is an important characteristic of a telescope. It is defined as the smallest angle between two close points which one can distinguish. Better angular resolution means smaller angle between the points.

For the circular objective the aperture – the width of the objective lense or mirror - is the diameter $\underline{D}$, and its area is equal to

$$P = \pi D^2/4 .$$

For monochromatic light with wavelength $\lambda$ the angular resolution $\underline{\delta}$ is given by the formula

$$\delta \approx 1{,}22\, \lambda\, /\, D$$

and the ratio $\delta/P$ is a significant quantity, too.

There is a clear trend: in order to gain better angular resolution very large telescopes are constructed. For instance the new VLT in California should have a diameter of 30 meters [1]. In those cases improving the resolution means increasing the diameter of the objective.

We would like to make a proposal aimed to improve the resolution without requiring gross enlargement of the objective element. It can be used for observation of objects which emit or reflect detectable amount of light, for example observation of objects on the Earth from satellites, observation of planetary systems in our Galaxy etc. The resulting lower weight is obviously an important advantage for the observation from satellites.

Telescope with a rotating objective element

In order to get better resolution we propose a new telescopic system involving a mount (paralactical, azimuthal or four axes) and an instrument for the digitalization of the image connected to a computer. The idea of this system is to use an objective element of suitable geometrical shape cut from the usual objective element in such a way that its projection on the plane perpendicular to the optical axis has one dominating dimension. The mount, however, has to perform one rotation more than the usual mounting: a rotation of the objective element or of the whole system around the optical axis is necessary.

For reflectors the new objective element has the form of a slice of the rotational paraboloid. The instrument for digitalization of images has to be located in the image plane of the telescope.

For refractors the new objective element is a part of circular lense with one dominating dimension and the instrument for digitalization of the image is located in the image plane of the telescope.

Of course, this idea can be used also for radiotelescopes.

Examples

      Examples of the new telescopic systems are presented in Figures 1 and 2. One possibility to create such telescopic system is to modify the Newton system. Here the objective mirror is a rectangular slice cut from a rotational paraboloid as shown in Fig. 1. The other Fig. 2 shows the details of an objective lense for the new telescopic system. These are the modifications we propose in order to improve the ratio resolution/area of common telescopic systems.

      The objective element has the shape of a rectangular slice of a circular objective element for the usual system. Projection of this slice on the plane perpendicular to the optical axis has dominating length $\underline{L}$ and width $\underline{B}$. The length determines the resulting resolution which is better for bigger length. In general the projection can have shapes like: rectangular, elliptic, part of a circle between two chords, a convex plane figure limited by arcs etc. As shown in Fig. 2, the dominating length $\underline{L}$ should be (much) greater than the width $\underline{B}$.

      Angular resolution of this system for monochromatic light with wavelength $\lambda$ is $\delta \approx \lambda / L = B\lambda / P$ in the direction of $\underline{L}$ and $\delta' \approx \lambda / B = L\lambda / P$ in the direction of $\underline{B}$. **L . B = P** is the area of the projection of objective. The ratio of both angular resolutions $\delta'/\delta$ then determines the number of images and the angles in which they have to be made and stored, in order to reconstruct the image with maximal resolution, i.e. with the resolution better than $\delta \approx \lambda / L$. The image is reconstructed by using simple software. The problem leads to numerical solution of a system of linear equations defined by each snap. For example, each cell of a CCD camera detects integral intensity of light coming from a rectangular part of a source with the length and width proportional to the length and width of the objective mirror. Using the mathematics of tomography can improve the resolution and the image.

      In some cases it is possible to use the rotation of the sky: it performs rotation of the objects with time according to the elevation, instead of rotation around the optical axis. The snaps have to be stored at different times given by the Earth rotation. It may also be advantageous to propose a system of „strip" telescopes or radiotelescopes from north to the south of the Earth, each of them observing the sky

near the zenith.  Strip means again that the length is many times greater than the width.The objective would be almost horizontal allowing simple construction. It would rotate around optical axis, i.e. in the almost horizontal plane. The increased resolution should help, at least, to find more planetary systems and more details of the sky. The lower weight is also useful for  launching telescopic systems to space for observations.

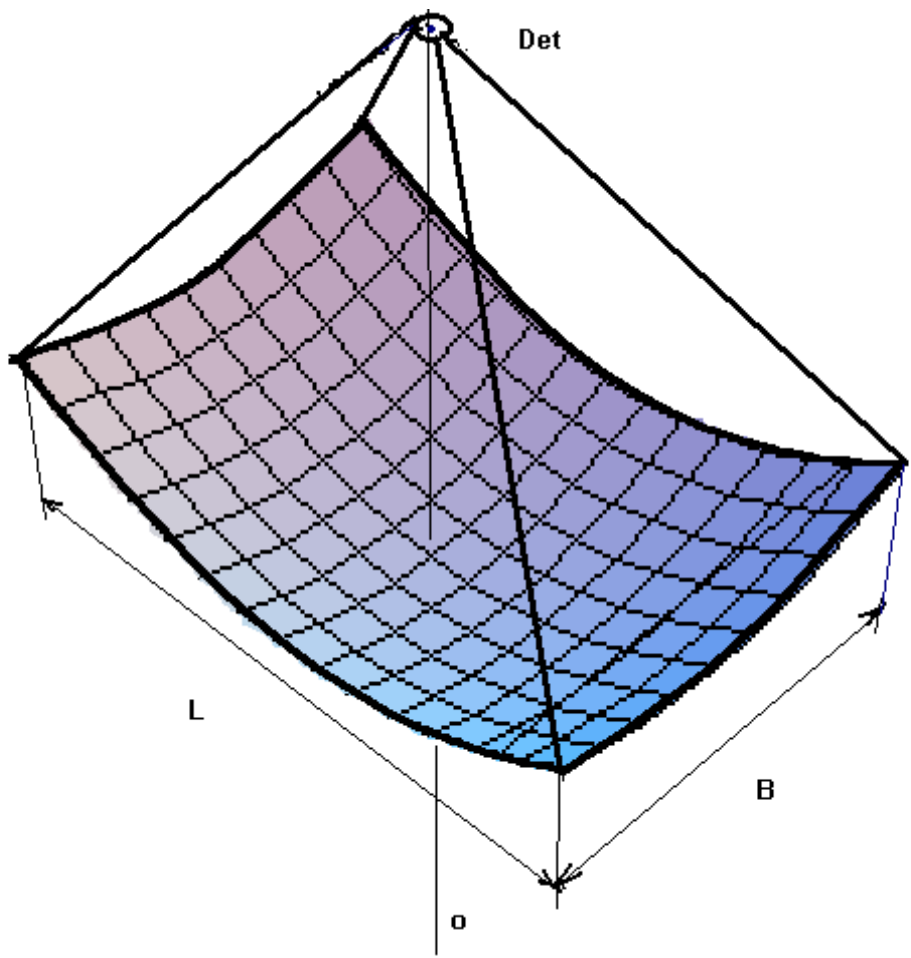

Fig. 1.   Principle of the new telescopic system. It consists of a parabolic objective mirror of rectangular shape and a CCD camera in the image plane. The objective mirror rotates around axis o and the shots are stored in a computer, where the image is reconstructed.

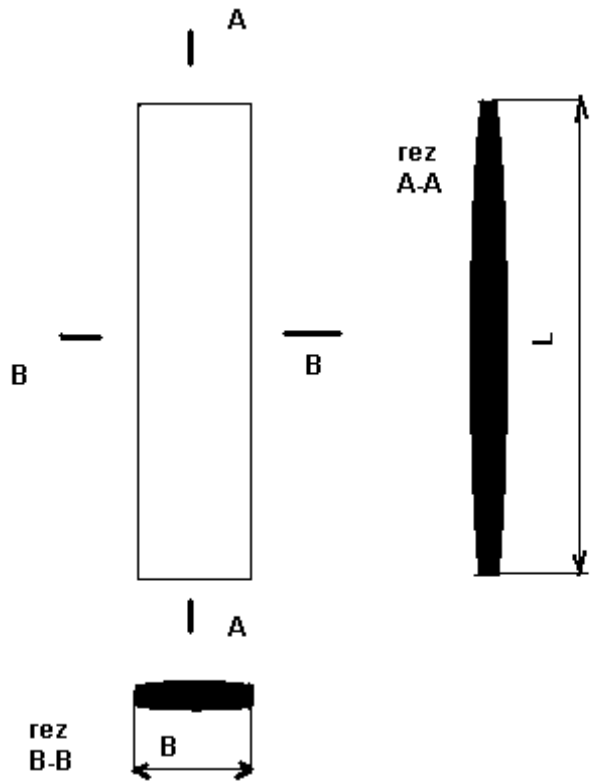

Fig. 2. Objective lense for a new telescopic system with rotational refractor.

References

[1] The Very Large Telescope Project,    http://www.eso.org/projects/vlt/

[2]  F.S. Crawford, Jr., *Berkeley Physics Course 3. Waves* (McGraw-Hill Book Co., New York  1968).